# Fundamental Limitations on Gain of THz QCL

L. D. Shvartsman and B. Laikhtman

The Racah Institute of Physics, The Hebrew University of Jerusalem, Jerusalem, 91904, Israel

*Abstract.* We analyze the main physical processes in quantum cascade lasers with spatial separation between the region of photon radiation and LO phonon emission providing depopulation of the lower level of the optical transition. Our purpose is to find reasons of reduction of the population inversion at low photon energy and search for ways to increase it. The expression for the population inversion is obtained from equation for simplified density matrix. This allows us, on one hand, to take into account coherence of tunneling between different levels and, on the other hand, to understand its role in transition probabilities in a simple way. We found out that the fundamental reason limiting population inversion in THz lasers in the energy uncertainty principle. The population inversion can be significantly increased by optimization of tunneling matrix element between the two regions and LO phonon emission time. The optimal value of the matrix element is smaller than its maximal possible value. As well, the optimal LO phonon emission time is larger than the time reached at LO phonon resonant emission.

## 1. Introduction

After the first fabrication of quantum cascade (QC) THz laser nearly two decades ago [1] there was a lot of activity in this field motivated by applications of THz radiation in biology[2,3], medicine[4,5], security[6] and spectroscopy[7-9]. Significant progress has been made in characteristics of THz QC lasers.[10-14] Frequency range is extended down to nearly 1 THz.[12-15] Improved injection scheme [16-18], oscillator strength [17] and thermoelectric cooling [19] made it possible to increase working temperature to 250 K[20]. Lasing power reached the watt level comparable to that of MIR QC lasers.[21-24] However, in spite of demand from applications and a lot of creativity of many groups the smaller frequency the smaller lasing power they succeeded to achieve. [25-28] And so far nobody succeeded to demonstrate lasing at 1THz.[16-18,29-31] Persistence of this limitation suggests that there are fundamental reasons behind it. The purpose of the present paper is to search for these reasons, and to study possible ways to improve laser characteristics.

To fulfil this purpose, we analyze physical mechanisms that provide power lasing. Trying to make our consideration as transparent as possible, we make use of a model and method that, on one hand, adequately describe these mechanisms and, on the other hand, is as simple as possible.

The main requirement necessary for powerful lasing is significant population inversion between the levels where the radiation transition takes place. So, we concentrate on the possibility of creation of the population inversion.

Our investigation shows that at small lasing energy maintenance of high population inversion is impossible. The reason is the following. To keep high population inversion it is necessary to empty the low level fast enough. However, short lifetime of electrons at the low level leads to their large energy uncertainty. At small lasing energy, i.e., at small energy separation between lasing levels this uncertainty becomes of the order or even larger than the energy separation. The levels strongly overlap leading to decrease of the population inversion. That is small width of the lasing levels requires long electron lifetime at the low level. But for appreciable population inversion this time has to be significantly smaller than non-radiative transition time between the upper and lower levels. This requirement limits the lifetime at the low level from above. The smaller lasing energy the stronger the limitations to the lifetime. That is the fundamental reason preventing maintenance of high population inversion at low lasing energy is the energy uncertainty principle.

The purpose of our paper is to make estimate of the limitation to lasing energy coming from the energy uncertainty principle for a realistic active region design. However, this estimate is sensitive to details of the design. Trying to emphasize generality of our approach and to make estimate as much universal as possible we make some simplifications and pretend only to order of magnitude validity of our results. More precise estimates have to take into account all details of the design and more precise values of parameters that we use. This can be done only separately for any particular structure. Such a task is beyond the scope of our paper.

There are two most popular kinds of the active region design of QC THz lasers: bound-to-continuum design [25,26,32-34] and resonant optical phonon design [35-43]. The contradiction between requirements of low lasing energy and short electron lifetime at the low level can be to some extent mitigated by increasing injection efficiency. This way is mainly used in bound-to-continuum design of the active region. Exactly in this design record low lasing frequency 1.2 THz was reached [17,18], however, with quite low lasing power (maximal cw power was 0.12mW at 10K). It is necessary to note that increase of the injection efficiency eventually leads to increase of the energy uncertainty of the upper level and also limits the decrease of the lasing energy.

To demonstrate the limitation put on the population inversion by contradicting requirements of short lifetime of the low lasing level and narrow width of this level we study this phenomenon in detail in resonant optical phonon design.

Our investigation is based on the observation that all variety of structures with different designs of the active region has the same system of levels in the active region that is critical for their work. We analyze a simple model that has this system of levels and concentrate on features and processes common for all structures with this system. In our investigation and conclusions we as much as possible avoid features that are substantially different for different structures and have to be investigated for each structure separately. Our purpose is to understand physical mechanisms limiting lasing power at low energies and to give order of magnitude estimate for this limitation but not to make more or less precise calculation for any specific model.

For MIR radiation typical is a three-level model that can be presented as three levels in a single well, Fig.1. (The structure may be more complicated, what is important for us is only the system of levels.) Electrons are pumped to level 2 with energy $\varepsilon_2$, go to level 1 with energy $\varepsilon_1$ emitting photons with energy $\hbar\omega = \varepsilon_2 - \varepsilon_1$ and then go to level 0 emitting LO phonon.[35,37,44,45] Population inversion between levels 1 and 2 is maintained by selectivity of LO phonon emission. The energy difference between levels 0 and 1, $\varepsilon_{10} = \varepsilon_1 - \varepsilon_0$ is very close or equal to LO phonon energy $\hbar\omega_{LO}$ that makes phonon emission time $\tau_{LO}(\varepsilon_{10})$ short and depletion of level 1 very effective. On the other hand, the energy difference between level 2 and level 0, $\varepsilon_{20} = \varepsilon_2 - \varepsilon_0$ is larger than $\hbar\omega_{LO}$ and transition from level 2 to level 0 with LO phonon emission requires large momentum transfer between electron and phonon. This decreases the transition matrix element increasing $\tau_{LO}(\varepsilon_{20})$ compared to $\tau_{LO}(\varepsilon_{10})$ and making depletion of level 2 ineffective and helping to maintain high population inversion.
.
However, with decrease of photon energy, $\hbar\omega = \varepsilon_2 - \varepsilon_1$ the difference between depletion time of level 2, $\tau_{LO}(\varepsilon_{20})$, and that of level 1, $\tau_{LO}(\varepsilon_{10})$, becomes small that makes it difficult to maintain population inversion between levels 1 and 2. Attempts to solve this problem by making the optical transition diagonal (i.e., states 0,1 and 2 in adjacent wells)[35], so as to reduce overlap 0,2 states have their disadvantages. Diagonal transition has small dipole matrix element and a broad emission linewidth due to interface roughness. To overcome this difficulty and provide high depletion selectivity it was suggested to separate photon radiation region and phonon emission region spatially.[38-43] The equivalent model of such design can be presented as a system of four levels in double quantum well, Fig.2. Electrons are pumped to level 3 in the left well, go to level 2 in the same well emitting photons, then tunnel to level 1 in the right well and finally go to level 0 in the right well emitting LO phonons. The objective of the spatial separation of photon radiation and phonon emission regions is to use resonant tunneling to ensure high depopulation selectivity. If level 2 is in resonance with level 1 while level 3 is not in resonance high depletion selectivity it was suggested to separate photon radiation region and phonon emission region spatially. [38-43] The equivalent model of such design can be presented as a system of four levels in double quantum well, Fig.2. Electrons are pumped to level 3 in the left then depletion time of level 2, $T_2$, is much shorter than depletion time of level 3, $T_3$. That is this approach is based on conception of two resonances: resonance between levels 1 and 2 that provides fast tunneling between them and resonance of the energy separation between levels 1 and 0 with LO phonon energy that helps to reduce phonon emission time from level 1. The basis for the difference between $T_2$ and $T_3$ is that the dependence of the tunneling probability between two levels on the energy separation between these levels is usually described by Lorentz contour that has its maximum when the energy separation is zero (see Fig.3 and Eqs.(3.2a) and (3.2b)). If levels 2 and 1 are in resonance, i.e. their energies are equal, while the energy separation between levels 2 and 3, $\varepsilon_{32} = \varepsilon_3 - \varepsilon_2$, is larger than the width of the resonance, i.e. the width of the Lorentz contour, then tunneling probability from level 2 to level 1 followed by phonon emission is much larger than the same process from level 3 even if $\tau_{LO}(\varepsilon_3) - \tau_{LO}(\varepsilon_2)$ is negligible.

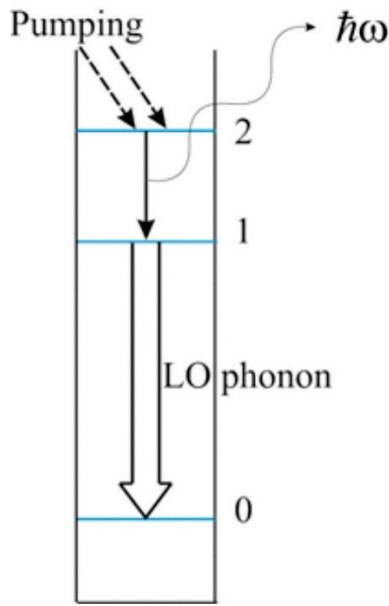

Fig.1. Typical system of levels for single well lasing.

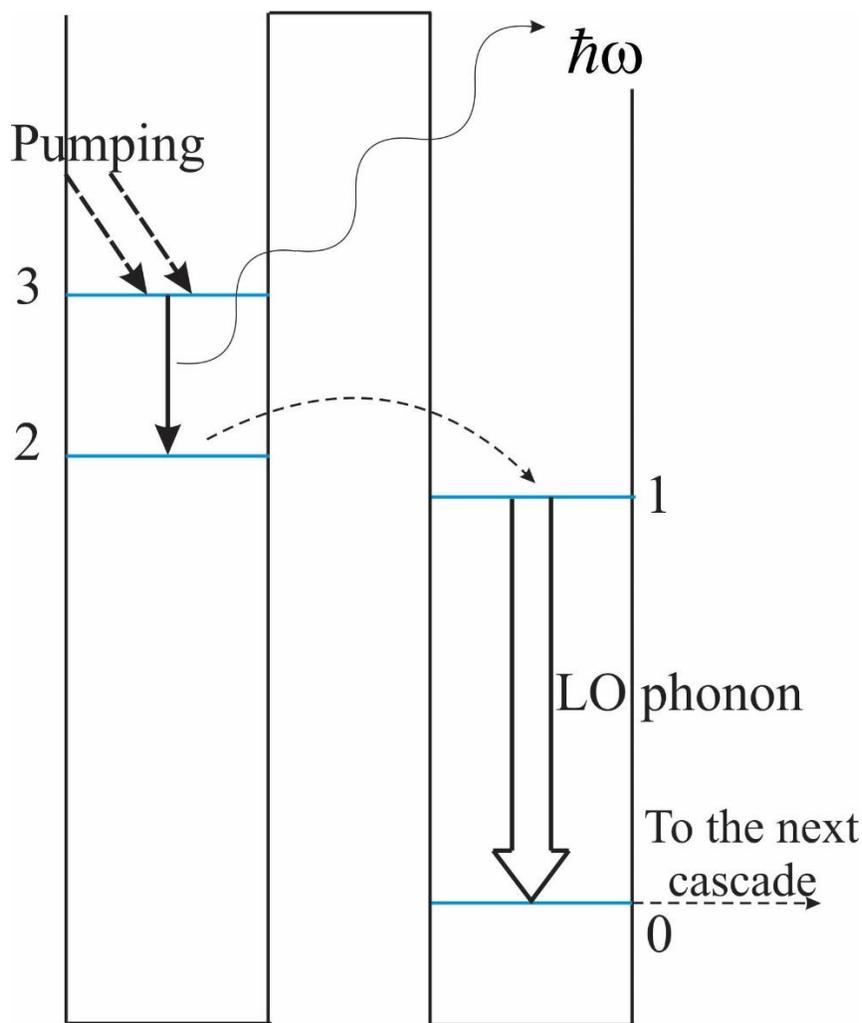

Fig.2. Typical system of levels with spatial separation photon and LO phonon emission regions.

In spite of the advantage of the structures with spatially separated radiation and LO phonon emission regions and fabrication of lasers working below 2 THz, lasing power reached in this region remains quite small. We demonstrate that reason of this limitation on the radiation energy is the energy uncertainty principle. We are also trying to show possible ways to weaken this limitation.

Maintenance of high population inversion at small radiation energy $\hbar\omega = \varepsilon_{32}$ requires the design of the structure with the width of level 2, i.e., the width the tunneling resonance much smaller than $\varepsilon_{32}$. Typically, the width of the resonance is controlled by energy uncertainty resulting from scattering, that in the case under consideration is LO phonon emission. Usually, it is not paid attention to the fact that coherent tunneling process also contributes to the width of the resonance. Its contribution is of the order of the energy of Rabi oscillations (i.e., Rabi frequency multiplied by Plank constant) that is tunneling matrix element. Therefore, the tunneling matrix element appears under two contradictory requirements. On one hand, it is desirable to make it as large as possible to reduce depletion time of level 2, $T_2$. The reason is that at large $T_2$ depletion of level 2 competes with influx of electrons to this level from level 3 due to non-radiative transitions that suppresses the population inversion. On the other hand, it is desirable to make it small to reduce the width of the resonance and improve depletion selectivity. The last requirement becomes especially pressing with reduction of $\hbar\omega = \varepsilon_{32}$. As a result, there exists an optimal value of the tunneling matrix element that provides the best possible population inversion.

Similar consideration is related to LO phonon emission time. We show that the optimal value of $\tau_{LO}$ leading to maximal possible population inversion is not necessary the shortest value of $\tau_{LO}$ corresponding to resonance between levels 1 and 2 in Fig.2. The same conclusion based on another argument was made in Ref. 46 (see also Ref. 23).

Our purpose is to estimate the dependence of maximal accessible population inversion between levels 2 and 3 on the energy separation between them. For this reason, we assume ideal injection selectivity, don't consider temperature [47] and other effects [48-50] that are detrimental to population inversion and gain. These effects make real population inversion even smaller than our estimate. Also these effects can be fought with improving technology, inventing a more favorable design, strengthening thermoelectric cooling and so on. On the other hand, fundamental physical limitations that we study cannot be fought or removed in any way.

To simplify our consideration we neglect $\varepsilon_{32}$ dependence of non-radiative transition time between levels 3 and 2. The precise value of the population inversion and a minimum of the lasing energy that we obtain for our simplified model is true in the order of magnitude. The exact values depend on details of real structure where it is used.

Trying to study population inversion at small energy separation between levels 2 and 3 we come to the situation where this separation becomes of the order of the width of the levels, i.e., the levels overlap. In this case it is difficult to say about electron transition $3 \rightarrow 2$ which of the states initial or final has higher energy. In this case it is more appropriate to consider gain directly that we shortly discuss in Sec.5.

In Sec.6 we study energy dependence of $\tau_{LO}$ near resonance of LO phonon emission, $\varepsilon_{10} = \hbar\omega_{LO}$ to find out how and to what extent energy separation $\varepsilon_{32}$ affects $\tau_{LO}(\varepsilon_2)$ and $\tau_{LO}(\varepsilon_3)$. We come to the conclusion that for realistic values of the non-radiative transition time optimization of the structure can allow to produce significant population inversion down the radiation energy around 1 THz. Finally, we are trying to point out possibilities to weaken this limitation.

All our calculations are made for the system of levels presented in Fig.2. Two wells shown there is just a simple equivalent model having these levels. Our results are applicable to any structure (i.e., three or more wells) with this system of levels.

## 2. Rate equations

To study population inversion at low energy of radiation and to produce reliable estimates we are not trying to give a precise consideration of all details of different real structures. Instead, we use a simple model that keeps most important features of all of them: spatially separated radiating region, phonon emission region and tunneling between them. This makes our consideration simple and clear.

Our model contains two wells separated by a barrier, Fig.2. Electrons are pumped to level 3 in the left well (radiating region) and go to level 2 radiating photons. From level 2 electrons tunnel to level 1 in the right well (phonon emission region) and go down to level 0 emitting LO phonons.

Simple and transparent description of the electron dynamics in the four-level system in Fig.2 is given by rate equations.[51-53] However, usually electron transport in rate equations is characterized by scattering times that don't include coherent processes and their dephasing. On the other hand, tunneling between different levels in cascade lasers is a coherent process. Many researches noticed that neglect of the coherence leads to substantial discrepancy between theoretical and experimental results and inclusion of coherence is necessary.[54-60] Complete description of electron transport in cascade lasers is provided with help of nonequilibrium Green function technique.[54,61-66] However, so detailed description needs heavy numerical calculations while accuracy and some details obtained with this technique compared to more simple approaches is excessive. E.g., renormalization of electron spectrum doesn't seem to be important for laser performance and hardly can be measured in practical devices. The approach adequate to the problem is simplified density matrix calculation that has been used by many researchers.[55-59, 67-72]. We also use this approach.

The Hamiltonian of the system

$$H = \begin{pmatrix} \varepsilon_3 & 0 & F_3 \\ 0 & \varepsilon_2 & F_2 \\ F_3 & F_2 & \varepsilon_1 \end{pmatrix} + H_{el-ph} + H_{32} \qquad (2.1)$$

includes the Hamiltonian of levels 1, 2, 3 shown in Fig. 2 Eq.(1.1) with energies $\varepsilon_1$, $\varepsilon_2$, $\varepsilon_3$ and tunneling matrix elements $F_2$ and $F_3$ between levels 2 and 3 and level 1 respectively, the Hamiltonian of LO phonons and their interaction with electrons $H_{el-ph}$ that provides electron transition between levels 1 and 0 with LO phonon emission, and electron scattering Hamiltonian, $H_{32}$, mainly electron – electron scattering, that provides non-radiative transitions between levels 3 and 2.

Dynamics of the system is described by Liouville equation for density matrix $\rho$

$$i\hbar \frac{\partial \rho}{\partial t} = [H, \rho] \qquad (2.2)$$

Matrix $\rho$ describes the whole system including electrons, phonons and other possible scatterers. To obtain the density matrix only for electrons it is necessary to average $\rho$ over states of all scatterers. The contribution of the two last terms in Hamiltonian (2.1) comes in only in the second order of perturbation theory.[72]

It is necessary also to keep in mind that electron states are characterized not only by level number 1, 2, 3 but also by electron momentum along the layer and spin. Without magnetic field in non-magnetic material different spin states are equally occupied and the spin variable can be discarded. To avoid unnecessary complication, we simplify the description assuming some fixed electron distribution (not necessary equilibrium) at each of the levels 1, 2 and 3 and discard the also electron momentum. However, tunneling between the levels is a coherent process that conserves in-layer electron momentum and some trace of this momentum remains. Electrons from level 3 with momentum **k** come to level 1 with the same momentum. But energies of the eigenstates at level 3 and level 1 with the same in-plane momentum are different. Therefore, although electrons coming to level 1 have the wave function of this level with momentum **k** their energy is different from eigenenergy of this level. In other words, this state is virtual, it is intermediate between the initial state at level 3 and tunneling back or emitting LO phonon.[67] Similar can be said about electrons tunneling between levels 2 and 1.* At level 1 electrons

*Footnote.* Note that for electrons at level 1 coming from level 2 this state is virtual even in case of resonance between levels 1 and 2. Without scattering an electron tunnels back and forth between these levels (Rabi oscillations) and can be localized at one of them only due to scattering. But the fastest scattering process is LO phonon emission that brings them to level 0.

coming from level 3 and 2 with the same *k* have different energies and electrons with the same energy have different momentum. That is these two groups of electrons are non-coherent, they don't mix and must be considered separately. For separate consideration in the density matrix formalism we have to introduce separate diagonal matrix element for each group. We call these matrix elements $\rho_{11}^{(2)}$ and $\rho_{11}^{(3)}$ where the subscript indicates the number of the level where electrons came from. As a result, we come up with the following equations for elements of the electron density matrix:

$$i\hbar \frac{\partial \rho_{33}}{\partial t} = F_3(\rho_{13} - \rho_{31}) - \frac{i\hbar}{\tau_{32}} \rho_{33} + i\hbar J \tag{2.3a}$$

$$i\hbar \frac{\partial \rho_{31}}{\partial t} = \varepsilon_{31} \rho_{31} + F_3(\rho_{11}^{(3)} - \rho_{33}) - \frac{i\hbar}{2\tau_{LO}^{(3)}} \rho_{31} \tag{2.3b}$$

$$i\hbar \frac{\partial \rho_{22}}{\partial t} = F_2(\rho_{12} - \rho_{21}) + \frac{i\hbar}{\tau_{32}} \rho_{33} \tag{2.3c}$$

$$i\hbar \frac{\partial \rho_{21}}{\partial t} = \varepsilon_{21} \rho_{21} + F_2(\rho_{11}^{(2)} - \rho_{22}) - \frac{i\hbar}{2\tau_{LO}^{(2)}} \rho_{21} \tag{2.3d}$$

$$i\hbar \frac{\partial \rho_{11}^{(2)}}{\partial t} = F_2(\rho_{21} - \rho_{12}) - \frac{i\hbar}{\tau_{LO}^{(2)}} \rho_{11}^{(2)} \tag{2.3e}$$

$$i\hbar \frac{\partial \rho_{11}^{(3)}}{\partial t} = F_3(\rho_{31} - \rho_{13}) - \frac{i\hbar}{\tau_{LO}^{(3)}} \rho_{11}^{(3)} \tag{2.3f}$$

$$\rho_{12} = \rho_{21}^*, \qquad \rho_{13} = \rho_{31}^* \tag{2.3g}$$

where $J$ is pumping, $\varepsilon_{31} = \varepsilon_3 - \varepsilon_1$, $\varepsilon_{21} = \varepsilon_2 - \varepsilon_1$, $\tau_{LO}^{(3)} = \tau_{LO}(\varepsilon_3)$ and $\tau_{LO}^{(2)} = \tau_{LO}(\varepsilon_2)$ is the optical phonon emission time of electrons coming to level 1 from levels 3 and 2 respectively. Our purpose is to estimate maximal possible population inversion. Therefore, considering non-radiative transition between levels 2 and 3 we take into account only transition from level 3 to level 2 characterized by time $\tau_{32}$ and neglect transition in the opposite direction assuming low enough temperature. We assume also that

$$\tau_{LO}^{(2)}, \tau_{LO}^{(3)} \ll \tau_{32} \tag{2.4}$$

and in Eq.(2.3) neglect the decay of off-diagonal matrix elements $\rho_{21}$ and $\rho_{31}$ coming from transitions between levels 3 and 2 compared to the decay coming from optical phonon emission.

Analytic results obtained directly from the equation for density matrix are very cumbersome [56-58,60,67] and to understand the role of each of the parameters it is necessary to analyze a lot of plots obtained numerically. We chose another way. We eliminate from Eq.(2.3) off-diagonal elements and obtain equations for only diagonal elements, i.e., rate equations. Relaxation and transition times that enter rate equations have simple physical meaning. They contain all information about coherence and dephasing and their dependence on original parameters of Eq.(2.3) is rather simple and tractable.

In steady state rate equations for populations of the levels $n_1^{(2)} = \rho_{11}^{(2)}$, $n_1^{(3)} = \rho_{11}^{(3)}$, $n_2 = \rho_{22}$ and $n_3 = \rho_{33}$ are

$$\frac{dn_3}{dt} = J - \frac{n_3}{\tau_{32}} + \frac{n_1^{(3)} - n_3}{T_{31}} = 0 \qquad (2.5a)$$

$$\frac{dn_1^{(3)}}{dt} = \frac{n_3 - n_1^{(3)}}{T_{31}} - \frac{n_1^{(3)}}{\tau_{LO}^{(3)}} = 0 \qquad (2.5b)$$

$$\frac{dn_2}{dt} = \frac{n_1^{(2)} - n_2}{T_{21}} + \frac{n_3}{\tau_{32}} = 0 \qquad (2.5c)$$

$$\frac{dn_1^{(2)}}{dt} = \frac{n_2 - n_1^{(2)}}{T_{21}} - \frac{n_1^{(2)}}{\tau_{LO}^{(2)}} = 0 \qquad (2.5d)$$

where the probability per unit time for an electron to tunnel to level 1 from level 2, $1/T_{21}$, or levels 3, $1/T_{31}$, and emit LO phonon are defined according to the relations

$$T_{21} = \frac{\tau_{LO}^{(2)}}{F_2^2}\left(\varepsilon_{21}^2 + \frac{\hbar^2}{4\tau_{LO}^{(2)2}}\right), \qquad T_{31} = \frac{\tau_{LO}^{(3)}}{F_3^2}\left(\varepsilon_{31}^2 + \frac{\hbar^2}{4\tau_{LO}^{(3)2}}\right) \qquad (2.6)$$

Their dependence on parameters is easy to understand. If an electron is initially at level 2 it appears at level 1 with probability $F_2^2/\varepsilon_{21}^2$. This is similar to two coupled classical oscillators: if oscillations of one of them are excited their energy is transferred to the second oscillator. In average the transferred energy is inverse proportional to the difference of frequencies of the oscillators squared. If the oscillations decay this decay leads to dephasing and renormalizes the oscillator frequencies. In quantum case the decay and dephasing results from energy uncertainty $\hbar/2\tau_{LO}$ coming from phonon emission. It destroys coherence of the state and localizes an electron in level 1. With this uncertainty the probability of tunneling from level 2 to level 1 is $F_2^2/(\varepsilon_{21}^2 + \hbar^2/4\tau_{LO}^{(2)2})$. An electron at level 1 goes to level 0 with LO phonon emission with probability per unit time $1/\tau_{LO}^{(2)}$. $1/T_{21}$ is the product of the tunneling and phonon emission probabilities. The structure of $1/T_{31}$ is similar.

### 3. Population inversion

Eq.(2.5) leads to the following well known expression for the population inversion [31,36,72-74]

$$\Delta n \equiv n_3 - n_2 = \left(1 - \frac{T_2}{\tau_{32}}\right) T_3 J \qquad (3.1)$$

where the probabilities per unit time to escape from levels 2 and 3 without ever coming back (i.e., depletion probabilities) are

$$\frac{1}{T_2} = \frac{1}{T_{21} + \tau_{LO}^{(2)}} = \frac{1}{\tau_{LO}^{(2)}} \frac{F_2^2}{\varepsilon_{21}^2 + F_2^2 + \hbar^2/4\tau_{LO}^{(2)2}} \qquad (3.2a)$$

$$\frac{1}{T_3} = \frac{1}{T_{31} + \tau_{LO}^{(3)}} + \frac{1}{\tau_{32}} = \frac{1}{\tau_{LO}^{(3)}} \frac{F_3^2}{\varepsilon_{31}^2 + F_3^2 + \hbar^2/4\tau_{LO}^{(3)2}} + \frac{1}{\tau_{32}} \qquad (3.2b)$$

The structure of these expressions is very simple. The probability to escape from level 2 per unit time is the product of the probability to tunnel from level 2 to level 1 multiplied by the probability of phonon emission per unit time. For electron at level 3 there are two escape channels. The first one is similar to that of level 2 and it is described by the first term in Eq.(3.2b). The second is non-radiative transitions to level 2 described by the second term in Eq.(3.2b).

Note that probability of coming from level 2 to level 1, $1/T_{21}$, differs from probability of escape from level 2 forever, $1/T_2$, by replacement of $\varepsilon_{21}^2$ with $\varepsilon_{21}^2 + F_2^2$. This happens because in the later case we allow for the possibility of electron tunneling back from level 1 to level 2 that leads to Rabi oscillations. Scattering localizes electrons at one of the levels destroying coherence of the Rabi oscillations. The same can be said about the difference between $1/T_{31}$ and $1/T_3$.

Typically $\tau_{LO} \sim 10^{-13} - 10^{-12}$ s [75], $\hbar/2\tau_{LO} \sim 3-0.3$ meV, $\tau_{32} \gtrsim 10^{-11}$ s [76,77] and $F_2 \approx F_3 \sim 3-7$ meV. At small $\varepsilon_{32}$ the difference between $F_3$ and $F_2$ is also small. In spite of exponential dependence of $F_3$ and $F_2$ on the barrier parameters, $F_3 \propto \exp\left[-\sqrt{2m(V-\varepsilon_3)}d/\hbar\right]$, $F_2 \propto \exp\left[-\sqrt{2m(V-\varepsilon_2)}d/\hbar\right]$, ($m$ is the electron effective mass under the barrier and $d$ is the barrier width) the height of the barrier $V - \varepsilon_3 \approx V - \varepsilon_2$ (several hundred meV) is so much larger than $\varepsilon_{32}$ that $(F_3 - F_2)/F_2 \approx \left(\sqrt{2m(V-\varepsilon_2)}d/\hbar\right)\left[\varepsilon_{32}/2(V-\varepsilon_2)\right]$ is small and for rough estimates the difference between $F_3$ and $F_2$ can be neglected. (We checked that if $F_3 > F_2$ by 15% $\Delta n/\tau_{32}J$ is smaller by about 3%.).

At small energy separation between levels 2 and 3, $\varepsilon_{32}$, the difference between $\tau_{LO}^{(3)}$ and $\tau_{LO}^{(2)}$ is also small and for rough estimates this difference can be neglected. We consider the effect of this difference at the end of Sec.4 and in Sec.5. In case when we neglect the difference between $F_3$ and $F_2$ or between $\tau_{LO}^{(3)}$ and $\tau_{LO}^{(2)}$ we will omit the number of the level.

Typical structure design corresponds to resonance between levels 2 and 1, $\varepsilon_{21}=0$, that increases the probability of tunneling, and facilitates depletion of level 2. In this case $\varepsilon_{31}=\varepsilon_{32}$. The second term in parentheses in Eq.(3.1) $T_2/\tau_{32} \approx \tau_{LO}^{(2)}/\tau_{32} \ll 1$ and can be neglected. When $\varepsilon_{32}$ is large the tunneling probability from level 3 to level 1 is suppressed so much that in spite of short $\tau_{LO}^{(3)}$ compared to $\tau_{32}$ the first term in Eq.(3.2b) is negligible compared to the second one and only one of two escape channels from level 3 remains. Then $T_3=\tau_{32}$ and population inversion reaches its maximal possible value,

$$(\Delta n)_{max} = \tau_{32} J \qquad (3.3)$$

This is the logic behind the conception of two resonances in MIR and FIR quantum cascade lasers.

However, when $\hbar\omega = \varepsilon_{32}$ decreases the first term in Eq.(3.2b) grows and becomes comparable with the second one. That is another depletion channel of level 3 opens up leading to decrease of $T_3$ and population inversion. This happens because of a finite width of the resonance tunneling followed by transition to level 0 with phonon emission, Fig. 3. Energy conservation requires that the energy difference between initial and final states of this process to be equal to LO phonon energy $\hbar\omega_{LO}$. However, finite width of the Lorentz contours in Fig. 3 shows that the process takes place in an interval of the energy separation between the initial and final state, i.e., even when it is different from $\hbar\omega_{LO}$. The possibility of such processes results from energy uncertainty of the initial state that comes from its final lifetime and shows up in the width of the Lorentz contour. Lorentz contour widths

$$\Gamma_2 = \sqrt{F_2^2 + \hbar^2/4\tau_{LO}^{(2)2}} \qquad \text{and} \qquad \Gamma_3 = \sqrt{F_3^2 + \hbar^2/4\tau_{LO}^{(3)2}} \qquad (3.4)$$

characterize the energy dependence of the transition probability from levels 2 and 3 to level 1 and then to level 0 and gives the order of magnitude of the energy uncertainty.

Fig. 3 presents a rough sketch of $1/T_2(\varepsilon_{21})$ and $1/T_3(\varepsilon_{31})$ that allows for their qualitative comparison. At the resonance between levels 2 and 1, $\varepsilon_{21}=0$ and depletion rate of level 2, $1/T_2$, reaches its maximum. Due to nonradiative transitions from level 3 to level 2 Lorentz contour $1/T_3(\varepsilon_{31})$ is lifted above $1/T_2(\varepsilon_{21})$ by the rate of these transitions $1/\tau_{32}$.

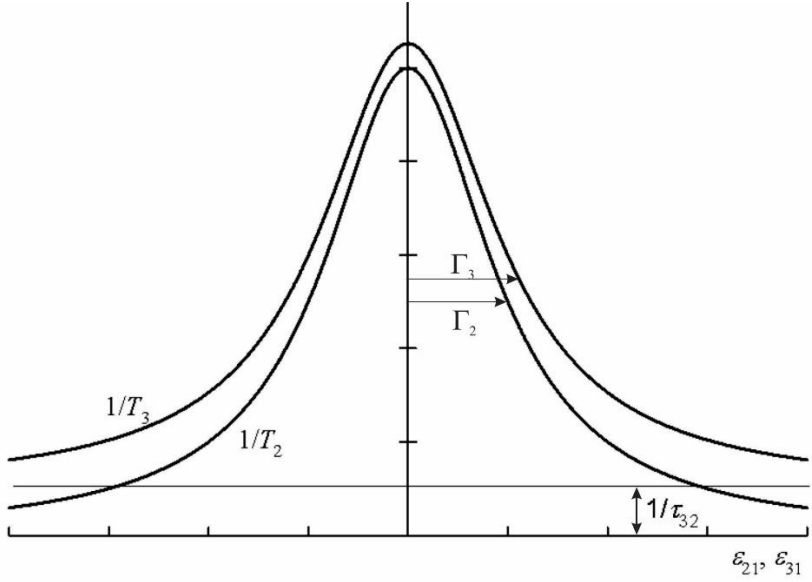

Fig.3. Sketch of $1/T_2(\varepsilon_{21})$ and $1/T_3(\varepsilon_{31})$. Both dependences are Lorentz contours with width $\Gamma_2$ and $\Gamma_3$ respectively, Eq.(3.4). Curve $1/T_3$ is lifted above $1/T_2$ by of non-radiative transitions rate from level 3 to level 2, $1/\tau_{32}$.

Since finite width of the tunneling resonance is critical for depletion of level 3 via tunneling followed by LO phonon emission one could think that this channel is negligible if $\varepsilon_{32}$ is kept much larger than the width of the resonance $\Gamma_3 \approx \Gamma_2$. However, this is not true because of high efficiency of this channel compared to non-radiative transitions due to fast phonon emission compared to non-radiative transitions between the levels, Eq.(2.4). Actually the first term in Eq.(3.2b) can be neglected compared to the second one only if $\varepsilon_{32}^2 + \Gamma_3^2 \gg F_3^2(\tau_{32}/\tau_{LO}^{(3)})$. Usually, the energy uncertainty due to phonon emission is not large compared to the tunneling matrix element, $\hbar/2\tau_{LO} \lesssim F$. Then this condition is reduced to

$$\varepsilon_{32}^2 \gg F_3^2\left(\frac{\tau_{32}}{\tau_{LO}^{(3)}}-1\right) \tag{3.5}$$

and the inequality has to be strong enough.

To reach maximal value of the population inversion one more condition is required along with Eq.(3.5). This condition is a very fast depletion of level 2 compared to non-radiative influx of electrons from level 3, i.e., $T_2 \ll \tau_{32}$ or

$$F_2^2 \gg \frac{\hbar^2}{4\tau_{LO}^{(2)2}\left(\tau_{32}/\tau_{LO}^{(2)}-1\right)} \tag{3.6}$$

This inequality is granted if $\hbar/2\tau_{LO} \lesssim F$ (the subscript of $F_3$ and $F_2$ is suppressed when the difference between them can be neglected) and Eq.(2.4) is true.
Inequalities (3.5) and (3.6) are necessary conditions to reach maximal value of the population inversion Eq.(3.3). The weaker the inequalities the smaller the maximal reachable population inversion.

## 4. Optimization of structure parameters

In this section we are trying to answer the question whether it is possible to keep high population inversion, and if yes then how, when $\hbar\omega = \varepsilon_{32}$ goes down and inequality (3.5) becomes weak.

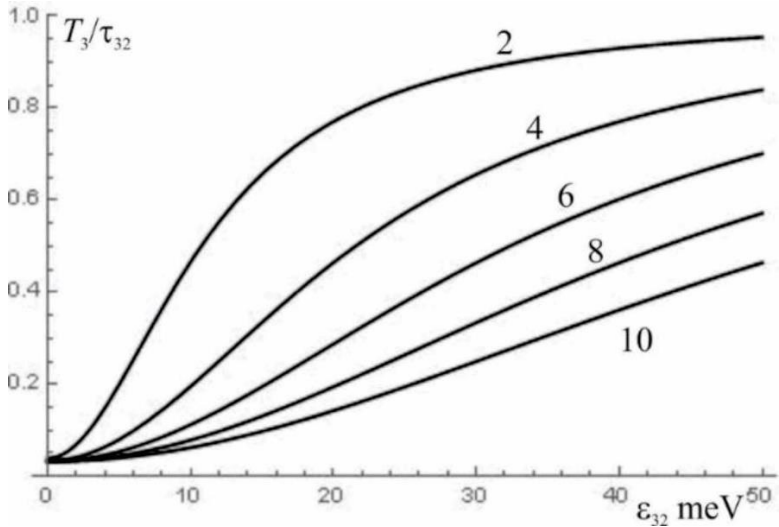

Fig.4. Dependence of the depletion time of level 3 on the energy separation between levels 3 and 2 for different values of the tunneling matrix element. Resonance between levels 1 and 2 is assumed, $\varepsilon_{21} = 0$. The plots are made for $\tau_{32} = 10$ps and $\tau_{LO}^{(3)} = 0.33$ps ($\hbar/2\tau_{LO}^{(3)} = 1$meV). Numbers near the curves are values of $F_3$ in meV.

Weakening of inequality (3.5) means strengthening of electron tunneling from level 3 to level 1. That is another depletion channel of level 3 opens up and $T_3$ falls off worsening the depopulation selectivity. The larger tunneling matrix element $F_3$ the larger is the value of energy separation $\varepsilon_{32}$ for this to happen, see Fig.4.
From this dependence one could conclude that long $T_3$ and high depopulation selectivity at small $\varepsilon_{32}$ can be reached by reduction of $F_3$ to keep it smaller than $\varepsilon_{32}$. However, this way has its strong limitation. $F_3 \approx F_2$ and reduction of $F_3$ by changing parameters of the barrier between the left and right wells in Fig.2 leads also to reduction of $F_2$. As a result, along with increase of $T_3$ increases also $T_2$ leading to weakening of inequality (3.6) and worsening depopulation of level

2. At very small $F_3 \approx F_2$ factor $(1-T_2/\tau_{32})$ in Eq.(3.1) becomes negative eliminating the population inversion at al. In Fig.3 this is illustrated by reducing the top of $1/T_2$ contour below $1/\tau_{32}$ line. Fig.5 shows the dependence of factors $(1-T_2/\tau_{32})$ and $T_3/\tau_{32}$ that control the population inversion, Eq.(3.1), on the tunneling matrix element for different values of $\varepsilon_{32}$ neglecting the difference between $F_3$ and $F_2$ (for this reason the subscript is suppressed). That is there are two contradictory requirements to $F$ (we suppress subscripts when the difference between $F_3$ and $F_2$ can be discarded): high population of level 3, i.e. $T_3$ as long as possible requires decrease of $F$ while high depopulation rate of level 2, i.e. $T_2$ as short as possible requires increase of $F$. These requirements are expressed in inequalities (3.5) and (3.6) that can be written as

$$\frac{\hbar/2\tau_{LO}^{(2)}}{\sqrt{\tau_{32}/\tau_{LO}^{(2)}-1}} \ll F \ll \frac{\varepsilon_{32}}{\sqrt{\tau_{32}/\tau_{LO}^{(3)}-1}}. \qquad (4.1)$$

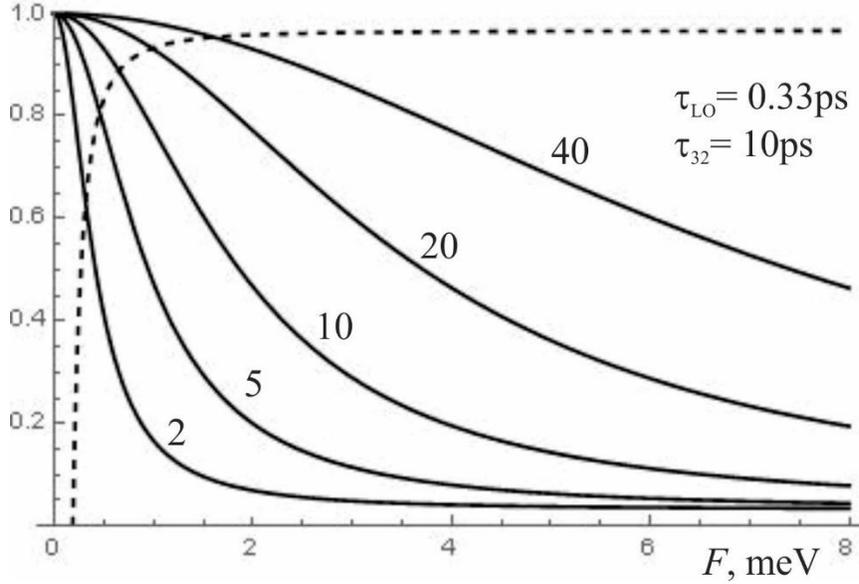

Fig. 5. Dependence of the depletion factors in Eq.(3.1) on the tunneling matrix element. Resonance between levels 1 and 2 is assumed, $\varepsilon_{21}=0$. Continuous lines show $T_3/\tau_{32}$ for different values of $\varepsilon_{32}$ (meV). Dashed line is $(1-T_2/\tau_{32})$.

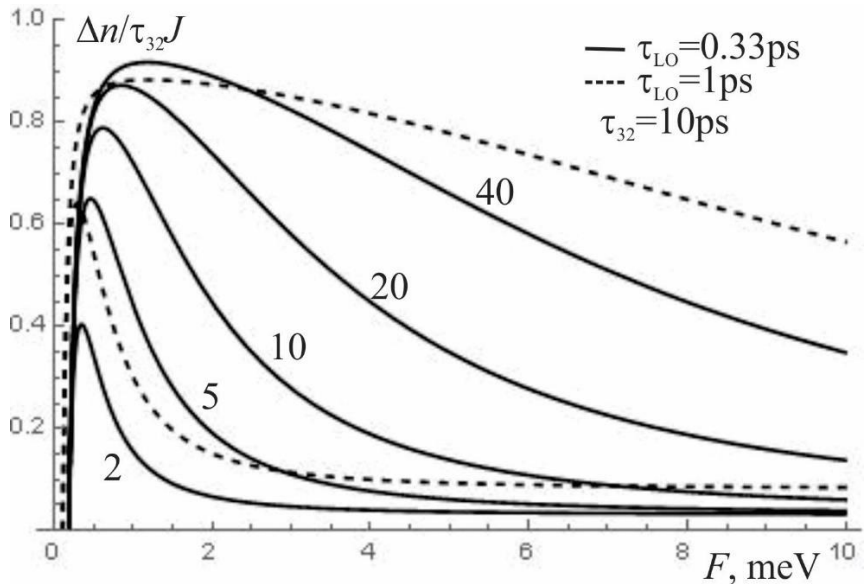

Fig. 6. Dependence of the population inversion on the tunneling matrix element. Resonance between levels 1 and 2 is assumed, $\varepsilon_{21}=0$. Numbers at the curves show values of $\varepsilon_{32}$ in meV. To see the effect of $\tau_{LO}$ variation plots for $\varepsilon_{32}=2\text{meV}$ and $\varepsilon_{32}=40\text{meV}$ are drawn for two different values of $\tau_{LO}$.

Two inequalities (4.1) lead to existence of an optimal value of $F$ that gives maximum to the population inversion, given $\varepsilon_{32}$. The exact expression for the optimal value of $F$ is very cumbersome but in leading order in $\hbar/2\tau_{LO}\varepsilon_{32} \ll 1$ it is

$$F^2 = \frac{\hbar}{2\tau_{32}} \frac{\varepsilon_{32}}{\sqrt{1-\tau_{LO}^{(2)}/\tau_{32}}} \sqrt{\frac{\tau_{LO}^{(3)}}{\tau_{LO}^{(2)}}}. \qquad (4.2)$$

With the same accuracy the population inversion corresponding to the optimal value of $F$ is

$$\Delta n = \left(1 - \frac{\tau_{LO}^{(2)}}{\tau_{32}}\right)\tau_{32}J. \qquad (4.3)$$

Asymptote (4.3) is reached only if optimization of $F$ is enough to keep both inequalities (4.1) strong. With reduction of $\varepsilon_{32}$ this becomes impossible and deviation from asymptotes (4.2) and (4.3) becomes quite noticeable especially for $\Delta n$. This is demonstrated in Fig.6 where dependence of the population inversion defined in Eq.(3.1) on $F$ for different values of $\varepsilon_{32}$ is shown and the difference between $\tau_{LO}^{(3)}$ and $\tau_{LO}^{(2)}$ is neglected.

As we already mentioned, the dependence of $\tau_{32}$ on $\varepsilon_{32}$ is neglected in all calculations. Our motivation is that this dependence is quite sensitive to details of the active region design while we are trying to make our estimates as universal as possible. This approximation leads to a loss of precision of our estimates, but it is not strong because in THz region $\varepsilon_{32}$ dependence of $\tau_{32}$ is weak. It is possible to see that this dependence would not change the character of curves in Fig.6 but lead their sharper variation of $\Delta n/\tau_{32}J$ with $F$.

Figs.(5) and (6) demonstrate two characteristic features:

(1) With decrease of $\varepsilon_{32}$ the decay of $T_3$ with growth of $F$ becomes steeper. As a result, the optimal value of $F$ becomes smaller and inequalities (4.1) weaken reducing the maximum value of $\Delta n$.

(2) Steeper decay of $T_3$ with growth of $F$ leads also to narrowing of the width of the maximum of $\Delta n$. Given $\varepsilon_{32}$, $\Delta n / \tau_{32} J > (\Delta n / \tau_{32} J)_{test}$ when the tunneling matrix element is within an interval of $\Delta F$ width. $\Delta F$ can be considered as a measure of the width of the maximum of the population inversion. The dependence of $\Delta F$ on $\varepsilon_{32}$ for several values of $(\Delta n / \tau_{32} J)_{test}$ is shown in Fig.7. The smaller the $\varepsilon_{32}$ the smaller $\Delta F$. $\Delta F$ goes to zero when the maximal value of $\Delta n / \tau_{32} J$ becomes lower than $(\Delta n / \tau_{32} J)_{test}$. Small values of $\Delta F$ makes fabrication of a structure with optimal value of $F$ at small $\varepsilon_{32}$ practically difficult.

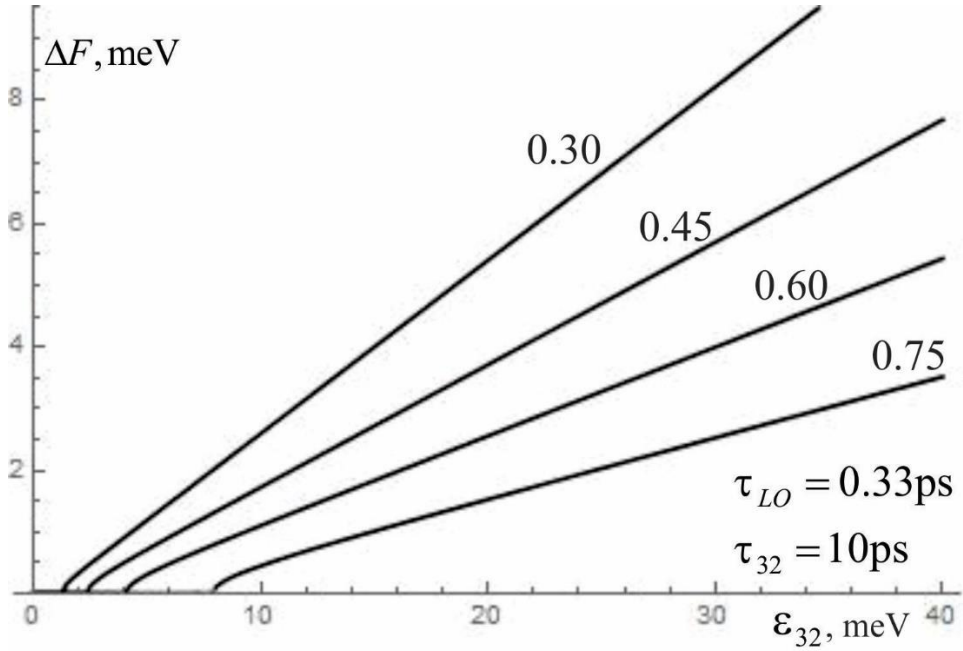

Fig.7. Dependence of the width of the maximum in Fig.6 on $\varepsilon_{32}$. To have $\Delta n / \tau_{32} J > (\Delta n / \tau_{32} J)_{test}$ the tunneling matrix element has to be within interval $\Delta F$. The curves present $\Delta F$ as a function of $\varepsilon_{32}$ for several values of $(\Delta n / \tau_{32} J)_{test}$ shown near the curves.

Two inequalities (4.1) are consistent under condition

$$\varepsilon_{32} \gg \frac{\hbar}{2\sqrt{\tau_{LO}^{(2)} \tau_{LO}^{(3)}}} \ . \tag{4.4}$$

The stronger inequality (4.4) the stronger inequalities (4.1) and the better results come from optimization of $F$. Weak inequality (4.4) means that the depletion rate of levels 2 and 3 via LO phonon emission becomes of the same `order and this makes it difficult to maintain high population inversion.

According to Eq.(4.4) one can expect that to make the population inversion larger with help of optimization of $F$ at low $\varepsilon_{32}$ it is necessary to increase $\tau_{LO}$ weakening inequality (2.4). This easily can be done if we take into account energy dependence of $\tau_{LO}$. $\tau_{LO}$ reaches its minimum at resonance, when the energy separation between levels 1 and 0 equals the phonon energy, see details in Sec.5. Violation of the resonance increases $\tau_{LO}$. Effect of this increase is shown in Fig.6 where two dashed curves are drawn for $\tau_{LO}$ larger than black curves. For now, we still neglect the difference between $\tau_{LO}^{(3)}$ and $\tau_{LO}^{(2)}$ that is not large at THz region of radiation and estimate the result of this difference later (see Fig.12).

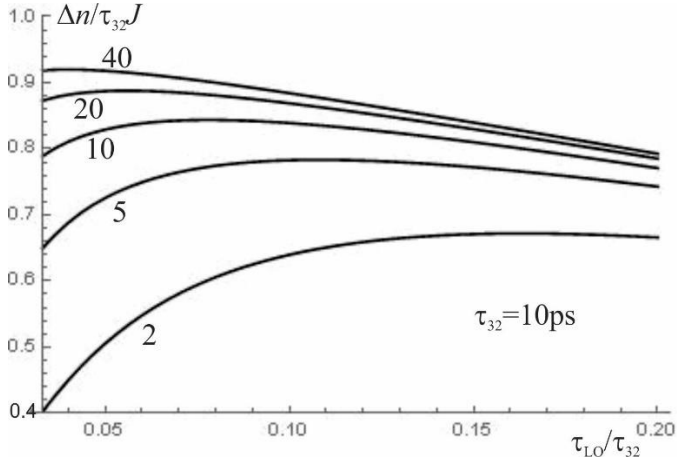

Fig. 8. Dependence of the population inversion reached at optimal value of $F$ on $\tau_{LO}$ Resonance between levels 1 and 2 is assumed, $\varepsilon_{21} = 0$. Numbers at the curves show values of $\varepsilon_{32}$ in meV.

But adjustment of the value of $\tau_{LO}$ is also under contradictory limitations. Increase of $\tau_{LO}$ makes inequality (4.4) stronger that leads to increase of $\Delta n$. However, $T_2 > \tau_{LO}$, therefore increase of $\tau_{LO}$ leads to increase of $T_2$, decrease of the first factor in Eq.(3.1) and drop of the population inversion. These contradictory limitations result in different effect of $\tau_{LO}$ increase at large and small values of $\varepsilon_{32}$. Fig. 6 demonstrates that at $\varepsilon_{32} = 40\text{meV}$ increase of $\tau_{LO}$ leads to decrease of maximal possible $\Delta n$ while at $\varepsilon_{32} = 2\text{meV}$ it makes possible to increase $\Delta n$. Dependence of $\Delta n$ reached at optimal value of $F$ on $\tau_{LO}$ for different values of $\varepsilon_{32}$ is shown in Fig. 8. As one can see the two contradictory requirements to $\tau_{LO}$ lead to existence of an optimal value of $\tau_{LO}$ that gives maximum to the population inversion. With growth of $\varepsilon_{32}$ dependence of the $\Delta n$ on $\tau_{LO}$ becomes more smooth and asymptotically the optimal value of $\tau_{LO}$ goes to zero (see Eq.(4.3):

the smaller $\tau_{LO}$ the larger $\Delta n$). Optimization of $\tau_{LO}$ becomes more important with decrease of $\varepsilon_{32}$ when $\Delta n$ falls off even at its optimal value.

The $\varepsilon_{32}$ dependence of maximal value of population inversion obtained as a result of optimization of both $F$ and $\tau_{LO}$ is shown in Fig. 9. This value is significantly larger than nonoptimized population inversion obtained for typical experimental values of $F = 3\text{meV}$, $\tau_{LO}$ =0.3ps and $\tau_{32} = 10\text{ps}$ (dashed line). The insert in Fig.9 shows the dependence of optimal ratio $\tau_{LO} / \tau_{32}$ dependence on $\varepsilon_{32}$.

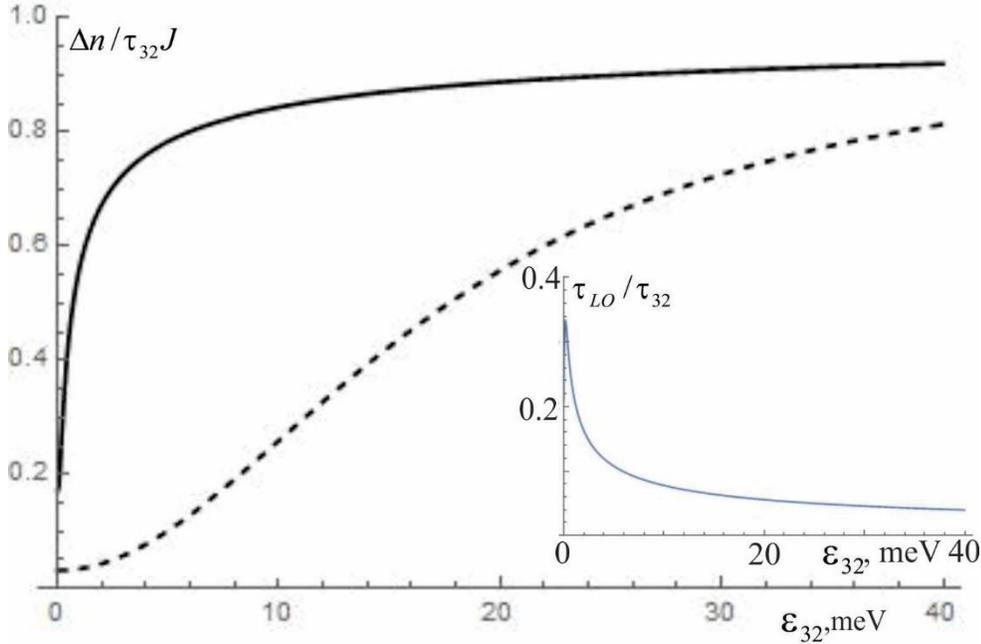

Fig. 9. Dependence of the population inversion on $\varepsilon_{32}$. Solid line is the population inversion reached by optimization of both tunneling matrix element and $\tau_{LO}$. Dashed line is non-optimized population inversion for $F$=3meV, $\tau_{LO}$=0.3ps and $\tau_{32} = 10\text{ps}$. The insert shows similar dependence of the optimal value of $\tau_{LO} / \tau_{32}$.

Increase of $T_2$ with $\tau_{LO}$ and weakening of inequality (2.4) shows that condition (4.4) has to be supplemented with one more condition that resulted in

$$\hbar\omega = \varepsilon_{32} \gg \frac{\hbar}{2\sqrt{\tau_{LO}^{(2)}\tau_{LO}^{(3)}}} \gg \frac{\hbar}{2\tau_{32}}. \tag{4.5}$$

Eq.(4.5) gives a fundamental limit for the region of $\varepsilon_{32}$ where it is possible to optimize population inversion. This limit is presented not as a threshold but as gradual decrease of the population inversion as inequalities (4.5) become weaker. The stronger these inequalities the higher population inversion can be created. The population inversion drops dramatically if the inequalities become weak.

Typically, $\tau_{LO}^{(2)} \approx \tau_{LO}^{(3)} \approx 0.3\text{ps}$ that gives $\hbar/2\tau_{LO} \approx 1\text{meV}$. If inequality (4.5) can be considered strong when the ratio of two different parts is more than 3 then it gives estimate $\varepsilon_{32} \approx 1\text{THz}$ for the minimal energy at which it is still possible to create significant population inversion.

## 5. Gain

So far in this paper we considered only population inversion that controls lasing power. However, strictly speaking lasing power is controlled by gain that is the product of population inversion and probability of photon emission. (As before we are interested in maximal reachable gain and discard losses and all detrimental effects.) The peak value of the gain at energy $\hbar\omega = \varepsilon_{32}$ is [74,75,78,79]

$$G = \frac{4\pi e^2 |z_{23}|^2 \varepsilon_{32}}{n_{eff} c\hbar L(\Gamma_2 + \Gamma_3)} \Delta n \tag{5.1}$$

where $L$ is the width of one cascade, $z_{23}$ is the dipole matrix element between initial and final lasing states and $n_{eff}$ is the effective refractive index at lasing frequency. The widths of the levels $\Gamma_2$ and $\Gamma_3$ depend on the tunneling matrix element and LO phonon emission time, Eq.(3.4). Therefore, optimization of the gain by appropriate choice of these parameters leads to values different from those optimal for $\Delta n$. Matrix element $z_{23}$ substantially depends on details of active region design. Its value and energy dependence can be significantly different for different structures. Therefore, the value and the energy dependence of the optimal gain cannot be estimated with the same degree of universality as $\Delta n$ and is beyond the scope of the present paper.

However, general conclusion that we made from investigation of the population inversion remains the same. The main physical limitation of the gain comes from the uncertainty principle. Decreasing energy separation between lasing level down to the width of these levels leads to their overlap that reduces depopulation selectivity and the gain. The energy uncertainty can be reduced by increasing of the lifetime of the low lasing level. But this reduction makes it comparable with characteristic time of nonradiative transition between the upper and lower levels that also reduces the population inversion. Given the energy separation between the levels the maximal possible gain can be reached by optimization of the depopulation of the levels that is controlled by the tunneling matrix element and LO phonon emission time.

## 6. Optical phonon emission. Beyond single $\tau_{LO}$

In this section we consider energy dependence of $\tau_{LO}$ and give more details of its effect on the population inversion. Our purpose is to find out how strong this dependence is at small $\varepsilon_{32}$ and to estimate to what extent it can improve the depopulation selectivity.
Assuming Boltzmann distribution of electrons at level 1 in the right quantum well in Fig.2, the electron transition rate from level 1 to level 0 for each phonon mode can be described by the following relation [80,81]

$$\frac{1}{\tau_{LO}} = \frac{e^2\sqrt{m}}{\sqrt{2\pi T}\hbar^2}\int_0^\infty e^{-(\Delta+E_q)^2/4E_qT}|M_{10}(q)|^2\,dq \qquad (6.1)$$

Where $\Delta = \hbar\omega_{LO} - \varepsilon_{10}$, $\omega_{LO}$ is the LO phonon frequency, $\varepsilon_{10} = \varepsilon_1 - \varepsilon_0$ is the energy separation between levels 1 and 0, $m$ is the electron effective mass, $T$ is electron temperature, $E_q = \hbar^2 q^2/2m$,

$$M_{10}(q) = \int \zeta_1(z)\zeta_0(z)f_q(z)dz \qquad (6.2)$$

is the matrix element of an effective electric charge distribution $f_q(z)$ created by LO phonon with in-plane wave vector $q$ between the initial $\zeta_1(z)$ and final $\zeta_0(z)$ electron wave functions in

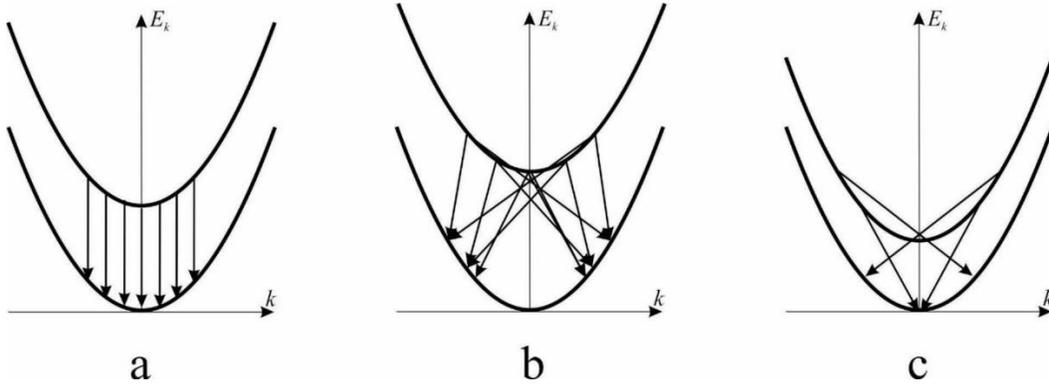

a          b          c

Fig. 10. Sketch of momentum and energy transfer from electrons to LO phonons in transitions from level 1 to level 0. Keep in mind that in reality electron in-plane wave vector and transferred wave vector $q$ are two-dimensional and have components perpendicular to the plane of the figure.

the quantum well. The dependence of $\tau_{LO}$ on $\varepsilon_{10}$ can be easily understood from Fig. 10 where the electron energy dependence on in-plane wave vector *k* at both 1 and 0 levels is shown along with scheme of transitions. If $\varepsilon_{10} = \hbar\omega_{LO}$ i.e. $\Delta = 0$, (Fig. 10a) then the transitions with zero momentum *q* transferred from electron to phonon are possible from all occupied states at level 1. If $\varepsilon_{10} > \hbar\omega_{LO}$, i.e. $\Delta < 0$, (Fig. 10b) then the transitions are possible also from all occupied states of level 1 but with nonzero momentum transfer that grows with difference $\varepsilon_{10} - \hbar\omega_{LO} = |\Delta|$ and is different for different initial states. The transition matrix element falls off with growth of the transferred momentum. As a result, the transition probability decreases. If $\varepsilon_{10} < \hbar\omega_{LO}$, i.e. $\Delta > 0$, (Fig. 10c) then the transitions from the bottom of level 1 are forbidden due to energy conservation. Transitions are possible only from states with high enough *k*. Occupation of these states falls off exponentially with energy and with it falls off the transition probability. Quantitatively this asymmetric behavior can be illustrated with asymptote of $1/\tau_{LO}$ at large values of $|\Delta|$:

$$\frac{1}{\tau_{LO}} = \frac{me^2}{2\hbar^2}|M_{10}(q_\Delta)|^2 e^{-(\Delta+|\Delta|)/2T} \qquad |\Delta| \gg T \qquad (6.3)$$

where $q_\Delta = \sqrt{2m|\Delta|}/\hbar$. Note that when *q* grows $|M_{10}(q)|^2$ falls off asymptotically as $1/q^2$.

Tunneling from the left well to the right well does not change electron energy and in-plane wave vector. Therefore, electrons from levels 2 and 3 come to level 1 virtually keeping their initial energy. Two resonance conception when levels 2 and 1 are in resonance and $\varepsilon_{10} = \hbar\omega_{LO}$ corresponds to Fig. 10a where for electrons at level 2 the probability to emit LO phonon is maximal and $\tau_{LO}$ is minimal. In this case $\tau_{LO}$ of electrons coming from level 3 can be calculated according to Eq.(5.1) where $\varepsilon_{10}$ should be replaced with $\varepsilon_{30} = \varepsilon_3 - \varepsilon_0$. This situation corresponds to Fig. 10b and with growth of $\varepsilon_{30}$ $\tau_{LO}$ grows but rather slowly, Fig.10.

Estimating the energy dependence of $\tau_{LO}$ quantitatively we have to keep in mind that each bulk mode of optical phonons is quantized in heterostructures [82-91]. If there is only one quantum well then there are phonons confined in the well, at each side of it and near the well interfaces. Frequencies of these phonons are different although not strongly. But the effective charge distributions created by each of them $f_q(z)$ are strongly different. This leads to different matrix elements Eq.(6.2) and eventually to different $\tau_{LO}$ for different modes and the same electron transition.[37] Dependence of their emission rate on $\varepsilon_{10}$ is shown in Fig.11. In more complicated structures the number of phonon modes is larger, and the splitting of their frequencies can also be larger. The first consequence of $\tau_{LO}$ dependence on energy is that $\tau_{LO}^{(3)} > \tau_{LO}^{(2)}$. This inequality means that actual depopulation selectivity is a bit better than its estimate obtained under assumption of $\tau_{LO}^{(3)} = \tau_{LO}^{(2)}$. To estimate how substantial this improvement is we compared the population inversion dependence on $F$ for $\varepsilon_{32} = 10\text{meV}$ (similar to one of the curves in Fig. 6) for two cases: one when the difference between $\tau_{LO}^{(2)}$ and $\tau_{LO}^{(3)}$ is neglected and the other when it is taken into account. In resonance when $\varepsilon_{21} = 0$ and $\varepsilon_{10} = \hbar\omega_{LO}$ Fig.11 gives $\tau_{LO}^{(2)} = 0.2\text{ps}$ and $\tau_{LO}^{(3)} = 2\tau_{LO}^{(2)}$. The two plots are shown in Fig.12. The plots show that at $\varepsilon_{32} = 10\text{meV}$ increase of $\tau_{LO}^{(3)}$ compared to $\tau_{LO}^{(2)}$ by two times leads to increase of the population inversion near its maximum by around 10%.

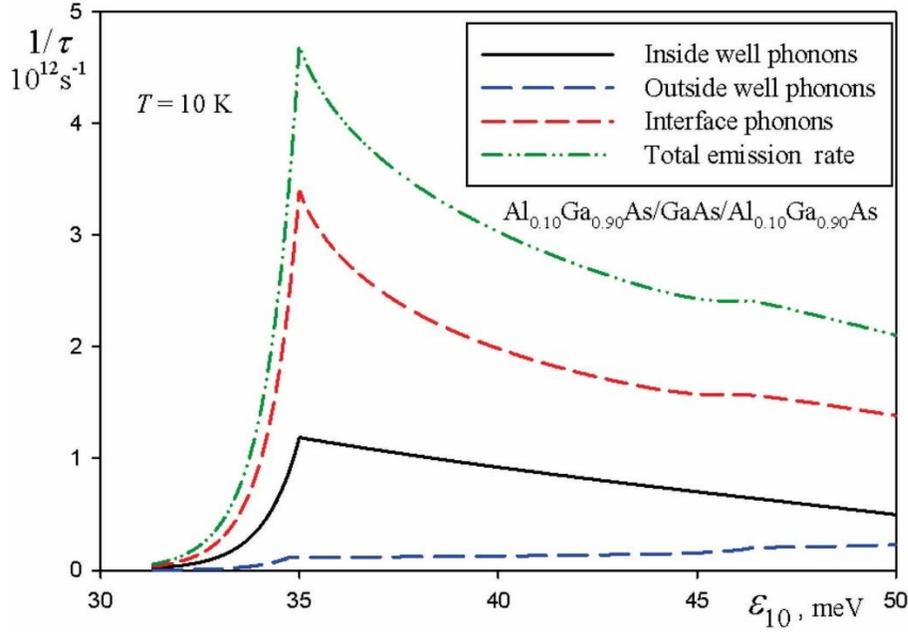

Fig.11. Emission rate of different phonon modes in a single quantum well.

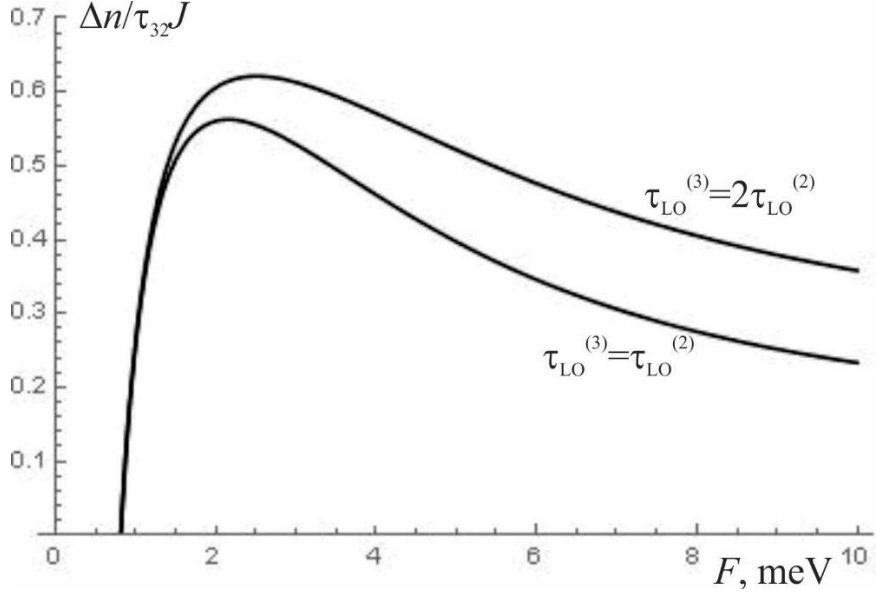

Fig.12. Comparison of the population inversion dependence on $F$ for equal and nonequal $\tau_{LO}^{(3)}$ and $\tau_{LO}^{(2)}$. In both cases $\tau_{LO}^{(2)} = 0.2\text{ps}$. For $\varepsilon_{32} = 10\text{meV}$ Fig.11 gives $\tau_{LO}^{(3)} = 2\tau_{LO}^{(2)}$. $\tau_{32} = 10\text{ps}$ is assumed.

The energy dependence of $\tau_{LO}$ can also be used for its increase to reach its optimal value, see Sec.4. This can be achieved by increasing $\varepsilon_{10}$. According to the character of the curves in Fig.10 $\tau_{LO}^{(2)}$ grows with energy faster than $\tau_{LO}^{(3)}$ and increase of $\varepsilon_{10}$ leads also to reduction of their difference.

We don't study effects the energy dependence of $\tau_{LO}$ in detail because the relation of such a study to experiment is not clear. The reason is that all calculations of this dependence are rather sensitive to the model. Typical practically used structures are chirped superlattices containing many interfaces. Quantization of optical phonons in such structures leads to quite a rich spectrum. Although the number of phonon degrees of freedom depends not on the number of interfaces but only on the number of unit cells, interfaces affect coupling between electrons and phonons.[37,89,92-96] E.g., outside modes for one well can be, at least partially, inside and interface modes for another well. Although this cannot lead to a very large increase of the emission rate because the electron wave function is smeared on several wells, its change by more than 10% is quite possible. This means that a reliable calculation of improvement of the population inversion due to the difference between $\tau_{LO}^{(3)}$ and $\tau_{LO}^{(2)}$ can be made only separately for any particular structure. As long as we are interested here in general, model independent results we don't study this effect in detail.

### 7. Conclusions.

In this paper we studied limitations of performance of QC lasers at low THz region. We assumed ideal injection selectivity and discarded temperature and other detrimental effects to estimate the highest possible population inversion that can be created at small lasing energy. We argue that even under these ideal conditions there exists a fundamental reason that prevents fabrication of powerful low energy cascade lasers. This reason is the energy uncertainty principle. Necessary condition for powerful lasing is high population inversion. Creation of high population inversion requires high depopulation selectivity and fast removal of electrons from lower lasing level. However, small electron lifetime at the low level increases electron energy uncertainty there and leads to large width of this level. At low lasing energy, i.e., small energy separation between the upper and lower levels this results in overlap between them. This overlap reduces the depopulation selectivity making high population inversion impossible. We demonstrate this on lasers with resonant LO phonon design of the active region.

Typical resonant LO phonon design of the cascade structure is based on conception of two resonances. First, tunneling between the low level of photon emission and upper level of LO phonon emission is made resonant to improve depopulation of the former level and, second, LO phonon emission is made resonant to reduce the phonon emission time and increase the depopulation rate. Our results show that population inversion reachable in this way drops with decrease of the lasing energy. In other words, we found that depopulation selectivity falls off when the photon energy becomes of the order of the width of the tunneling resonance that essentially is the energy uncertainty of the low level.

Maintenance of high population inversion applies contradictory requirements to tunneling matrix element and rate of the phonon emission. On one hand, they have to be small to ensure narrow tunneling resonance and, on the other hand, they have to be large to provide fast depletion of the low level of the optical transition. Decrease of the radiating photon energy exacerbates this contradiction leading to decrease of the population inversion. This is the main reason of power limitation of THz QC lasers.

We show that given lasing energy the population inversion can be increased by optimization of tunneling matrix element and phonon emission rate. The optimal value of the tunneling matrix element decreases with decrease of the lasing energy. The optimal value of the LO phonon emission time is larger than the value reached at the phonon emission resonance. That is the optimization leads to increase of the electron lifetime at the low lasing level mitigating the limitation applied by the uncertainty principle and means refusal of the conception of two resonances. But the inevitable increase of the optimal electron lifetime at the low level with decrease of the lasing energy still leads to reduction of the population inversion and drop of the lasing power. In other words, the optimization does not prevent the decrease of the laser power with the lasing energy, it can just significantly weaken it.

The optimization helps only if the energy of emitted photons is much larger than the energy uncertainty, Eq.(4.5). So, Eq.(4.5) present an inviolable condition put to the lasing energy by the uncertainty principle. Our estimates show that this fundamental limit of the radiation frequency is around 1 THz. Lasing energy dependence on maximal reachable population inversion is shown in Fig.8.

A natural question is whether it is possible to weaken limitation (4.5). One possibility is to modify the structure design making the wave function of level 3 shifted to the left of levels 1 and 2. Such a design would make non-radiative transitions $3 \rightarrow 2$ diagonal, decreases $F_3$ and increases both $\tau_{32}$ and $\tau_{LO}^{(3)}$. This design also could improve pumping selectivity. Typically it is used without tunneling and with LO phonon emission from low lasing level.[97-99] A well known disadvantage of this design is that optical transition is also diagonal reducing optical matrix element $z_{23}$. That is in this design there is a competition of two effects that affect gain in the opposite way: increase of the population inversion that increases the gain and decrease of the optical matrix element that decreases the gain. It makes sense also to consider alternatives [15,100-102].

## Figure captions

Fig.1. Typical system of levels for single well lasing.

Fig.2. Typical system of levels with spatial separation photon and LO phonon emission regions.

Fig.3. Sketch of $1/T_2(\varepsilon_{21})$ and $1/T_3(\varepsilon_{31})$. Both dependences are Lorentz contours with width $\Gamma_2$ and $\Gamma_3$ respectively, Eq.(3.4). Curve $1/T_3$ is lifted above $1/T_2$ by of non-radiative transitions rate from level 3 to level 2, $1/\tau_{32}$.

Fig.4. Dependence of the depletion time of level 3 on the energy separation between levels 3 and 2 for different values of the tunneling matrix element. Resonance between levels 1 and 2 is assumed, $\varepsilon_{21}=0$. The plots are made for $\tau_{32}=10$ps and $\tau_{LO}^{(3)}=0.33$ps ($\hbar/2\tau_{LO}^{(3)}=1$meV). Numbers near the curves are values of $F_3$ in meV.

Fig. 5. Dependence of the depletion factors in Eq.(3.1) on the tunneling matrix element. Resonance between levels 1 and 2 is assumed, $\varepsilon_{21}=0$. Continuous lines show $T_3/\tau_{32}$ for different values of $\varepsilon_{32}$ (meV). Dashed line is $(1-T_2/\tau_{32})$.

Fig. 6. Dependence of the population inversion on the tunneling matrix element. Resonance between levels 1 and 2 is assumed, $\varepsilon_{21}=0$. Numbers at the curves show values of $\varepsilon_{32}$ in meV. To see the effect of $\tau_{LO}$ variation plots for $\varepsilon_{32}=2$meV and $\varepsilon_{32}=40$meV are drawn for two different values of $\tau_{LO}$.

Fig.7. Dependence of the width of the maximum in Fig.6 on $\varepsilon_{32}$. To have $\Delta n/\tau_{32}J$ larger than some value the tunneling matrix element has to be within interval $\Delta F$. The curves show $\Delta F$ as a function of $\varepsilon_{32}$ for several values of $\Delta n/\tau_{32}J$.

Fig. 8. Dependence of the population inversion reached at optimal value of $F$ on $\tau_{LO}$ Resonance between levels 1 and 2 is assumed, $\varepsilon_{21}=0$. Numbers at the curves show values of $\varepsilon_{32}$ in meV.

Fig. 9. Dependence of the population inversion on $\varepsilon_{32}$. Solid line is the population inversion reached by optimization of both tunneling matrix element and $\tau_{LO}$. Dashed line is non-optimized population inversion for $F=3$meV, $\tau_{LO}=0.3$ps and $\tau_{32}=10$ps. The insert shows similar dependence of the optimal value of $\tau_{LO}/\tau_{32}$.

Fig. 10. Sketch of momentum and energy transfer from electrons to LO phonons in transitions from level 1 to level 0. Keep in mind that in reality electron in-plane wave vector and transferred wave vector $q$ are two-dimensional and have components perpendicular to the plane of the figure.

Fig.11. Emission rate of different phonon modes in a single quantum well.

Fig.12. Comparison of the population inversion dependence on $F$ for equal and nonequal $\tau_{LO}^{(3)}$ and $\tau_{LO}^{(2)}$. In both cases $\tau_{LO}^{(2)} = 0.2\text{ps}$. For $\varepsilon_{32} = 10\text{meV}$ Fig.11 gives $\tau_{LO}^{(3)} = 2\tau_{LO}^{(2)}$. $\tau_{32} = 10\text{ps}$ is assumed.